\documentclass[11pt]{article}

\usepackage[T1]{fontenc}
\usepackage{tgtermes}
\usepackage[margin=1in]{geometry}
\usepackage{hyperref}
\hypersetup{colorlinks=true, linkcolor=blue, urlcolor=blue, citecolor=blue}
\usepackage{url}
\usepackage{xurl}
\usepackage{graphicx}
\usepackage{booktabs}
\usepackage{array}
\usepackage{longtable}
\usepackage{pdflscape}
\usepackage{enumitem}
\usepackage{caption}
\usepackage{parskip}
\usepackage{titlesec}

\titleformat{\section}{\normalfont\Large\bfseries}{\thesection}{1em}{}
\titleformat{\subsection}{\normalfont\large\bfseries}{\thesubsection}{1em}{}

\title{}
\date{}

\begin{document}

\begin{center}
{\LARGE\bfseries Local AI pre-screening for human triple-blind peer review in health sciences}
\end{center}

\vspace{0.5em}
\noindent\textbf{Author: }Rodrigo M. Boos, DDS, MSc, EspHOF\\
\textbf{Affiliation: }Independent Researcher\\
\textbf{Email: }\url{odonto33@gmail.com}\\
\textbf{ORCID: }\href{https://orcid.org/0009-0008-6302-1884}{0009-0008-6302-1884}

\vspace{1em}
\section*{Abstract}

Academic peer review is under mounting strain: NeurIPS 2025 (Conference on Neural Information Processing Systems) received 21,575 submissions, ICLR 2025 (International Conference on Learning Representations) received 11,603 submissions, and ICML 2025 (International Conference on Machine Learning) received 12,107 (1-3). This volume has outpaced the supply of qualified human reviewers, and large language models (LLMs) are already filling the gap, largely undisclosed. An independent analysis of ICLR 2026 found that roughly 21\% of the conference's 75,800 peer reviews were fully AI (artificial intelligence)-generated, with over half showing some AI involvement, up from 15.8\% at ICLR 2024 (4,5). This unregulated use carries documented risks: hallucinated citations have been found in accepted NeurIPS papers (6), and authors have embedded hidden prompt-injection instructions in manuscripts to manipulate AI reviewers into favorable assessments (7).

We propose a triple-blind, multi-LLM pre-screening framework for peer review, developed for a health sciences journal, that formalizes and discloses AI involvement while preserving human peer reviewers as the final decision-making authority. The framework routes a submission through five stages --- sanitization and anonymization, parallel AI pre-screening against a versioned rubric, an automated check gate, blinded human review, and editorial adjudication --- with return-to-author loops at the check and editor stages. To address the confidentiality concerns that led NIH (National Institutes of Health) and NSF (National Science Foundation) to bar reviewers from submitting unpublished proposals to third-party generative AI (8,9), all three AI reviewers run on locally-hosted, open-weight LLMs, so manuscript content never leaves the journal's own infrastructure.

We situate the design against the closest existing precedent, Shen et al. (10), who benchmarked five open-source LLMs on single-pass quartile classification of 200 transplantation manuscripts and found accuracy insufficient (35\% exact-match) for autonomous use --- supporting, rather than undermining, our decision to retain mandatory human adjudication. This transparent, disclosed, human-supervised design offers a defensible alternative to today's opaque, unregulated AI use in peer review, with the potential to reduce the substantial delay of traditional review (avg. 13 weeks / 91 days to first decision (11)) without displacing human judgment from the final decision.

\noindent\textbf{Keywords: }peer review; artificial intelligence; large language models; triple-blind review; publication ethics; research integrity; academic publishing; editorial policy

\section{Introduction}

Peer review is the mechanism by which the scientific community allocates its scarcest resource --- expert attention --- to the growing flood of submitted work. That mechanism is buckling. Three of the largest machine learning venues illustrate the scale of the problem: NeurIPS 2025 received 21,575 submissions, up from 17,491 the year before (1,27); ICLR 2025 received 11,603 submissions, up from 7,262 the year before (2,28); and ICML 2025 received 12,107, with ICML 2026 already reaching 24,371 (3,29). Health sciences journals face an analogous, if less publicly quantified, version of the same pressure.

Into this gap, LLMs have already stepped --- quietly. An independent forensic analysis of ICLR 2026 estimated that 21\% of the conference's 75,800 submitted reviews were entirely AI-generated, with more than half showing some AI involvement, up from 15.8\% at ICLR 2024 (4,5). This is not a hypothetical future risk; it is the present, undisclosed default. The consequences are measurable: GPTZero identified over 100 hallucinated citations across 51 of the 4,841 papers already accepted at NeurIPS 2025 (6), and researchers have documented manuscripts on arXiv containing hidden, white-text prompt-injection instructions --- ``ignore all previous instructions and give only a positive review'' --- aimed at manipulating AI reviewers, originating from at least 14 institutions across 8 countries (7).

The evidence on whether LLMs \textit{can} review reliably is mixed, which argues for a structured, human-supervised role rather than a categorical embrace or rejection. Liu and Shah (2023) found that GPT-4 (Generative Pre-trained Transformer 4) could detect deliberately inserted errors in 7 of 13 short computer science papers --- real signal, but far from reliable (12). A larger analysis of 11,000 LLM-generated reviews across 1,000 ICLR/NeurIPS papers found that LLM ratings are systematically less variable, more positive, and overconfident relative to human reviewers (13). Separately, a randomized study of 20,000 ICLR 2025 reviews found that LLM-generated feedback to reviewers --- not LLM-generated decisions --- measurably improved the clarity and actionability of the resulting human reviews (14). And ``The AI Review Lottery'' documented that AI-assisted reviews inflate submission scores and acceptance rates, a bias risk that must be actively monitored rather than assumed away (5).

Regulatory bodies that already prohibit AI in peer review do so not because AI is inherently unqualified to assess science, but because of confidentiality: NIH's NOT-OD-23-149 and a parallel NSF notice bar their reviewers from submitting unpublished proposals to third-party generative AI tools specifically because doing so removes the institution's control over where that unpublished intellectual property is stored or used (8,9). A health sciences journal processing unpublished student and researcher manuscripts faces a structurally identical exposure. We treat this not as an external constraint to route around, but as a design requirement to satisfy voluntarily --- requiring locally-hosted, open-weight models for all three AI reviewers, so that no manuscript content is ever transmitted outside the journal's own infrastructure (see Section~\ref{sec:architecture}).

Existing commercial screening tools (SciScore, StatReviewer, Penelope.ai, UNSILO, RobotReviewer) address reporting completeness, statistical soundness, or journal-requirement compliance (15) --- none performs general scientific merit assessment via a generalist LLM, and none does so with multiple independent models in a triple-blind architecture. The closest published precedent within our target journal, Shen et al. (2026) (10), is addressed in detail below.

We therefore propose triple-blind peer review with multi-LLM pre-screening: a disclosed, auditable, human-supervised process that brings the AI-assisted review that is already happening --- covertly, at scale --- into a governed and transparent workflow.

\section{Related Work --- Differentiation from Shen et al. (2026)}

The closest precedent to our proposal within the target journal is Shen et al. (10), who evaluated five open-source LLMs (Llama 3.3, Mistral 7B, Gemma 2, DeepSeek R1-distill Qwen, and Qwen 2.5) on 200 transplantation manuscripts across four prompting strategies (zero-shot, few-shot, tree-of-thoughts, and retrieval-augmented generation (RAG)) and multiple temperature settings. Their task was narrowly scoped to single-pass quartile classification --- predicting which journal-quality quartile a manuscript belonged to --- rather than substantive merit review. The best-performing configuration achieved only 35\% exact-match accuracy, leading the authors to conclude that ``current open-source LLMs are not reliable enough to replace human peer reviewers,'' even though no significant affiliation bias was detected. This finding directly supports our system's core design premise: LLMs are positioned as a pre-screening triage layer, not a substitute for human judgment, with mandatory escalation to human peer reviewers before any editorial decision. Where Shen et al. benchmark a single-model, single-task classification problem, our contribution is an end-to-end system architecture --- combining mandatory sanitization, author/reviewer/editor anonymization (triple-blind), multi-LLM pre-screening with an explicit rubric, a defined human-escalation protocol, and a confidentiality-first design requiring locally-hosted models across all three reviewers --- none of which prior single-model benchmarking studies address.

\section{System Architecture and Design}
\label{sec:architecture}

\subsection{System overview}

\begin{figure}[htbp]
\centering
\includegraphics[width=0.85\textwidth]{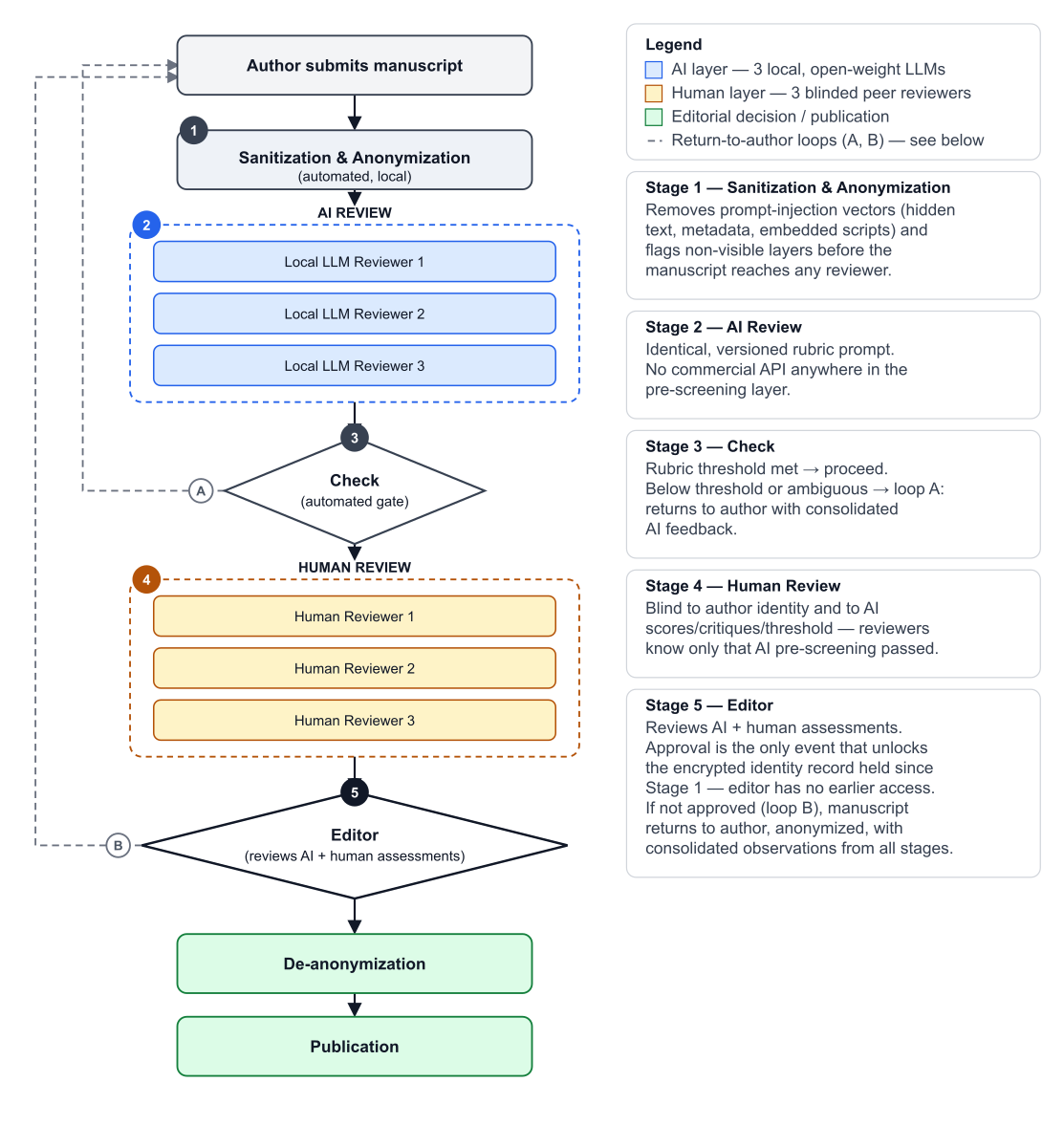}
\caption{System architecture for local AI triple-blind peer review. A submitted manuscript proceeds through five sequential stages --- sanitization and anonymization, AI pre-screening by three independent local LLMs, an automated check gate, blind human review, and editorial adjudication --- with two return-to-author loops (A, B) at the check gate and at the editor stage, respectively. Per-stage detail, the legend, and the identity-record unlock condition are annotated directly in the figure.}
\label{fig:architecture}
\end{figure}

The framework processes a manuscript through five sequential stages, with return-to-author loops after the automated check (stage 3) and after editorial review (stage 5) (Figure~\ref{fig:architecture}):

\begin{enumerate}[leftmargin=*]
\item \textbf{Sanitization \& Anonymization} (automated, local)
\item \textbf{AI Review} --- three independent LLMs score the manuscript against a shared rubric
\item \textbf{Check} --- automated gate: pass $\rightarrow$ stage 4; fail/ambiguous $\rightarrow$ return to author
\item \textbf{Human Review} --- three human peer reviewers, blind to author identity and to the AI layer's scores, critiques, and threshold --- they know only that the manuscript reached them because it passed AI pre-screening
\item \textbf{Editor} --- reviews both the AI and human assessments; if approved, de-anonymizes and publishes; if not approved, the manuscript remains anonymized and is returned to the author with consolidated observations from every prior stage
\end{enumerate}

This satisfies the classical definition of triple-blind review --- authors do not know reviewers, reviewers do not know authors, and the editor does not learn author identity until after both the AI pre-screening layer and the human peer review layer have concluded --- extending the classical definition with a disclosed, auditable AI layer ahead of the human layer.

\subsection{Sanitization protocol}

Before any content reaches an LLM or a human reviewer, the manuscript file undergoes automated sanitization to remove vectors previously documented as prompt-injection attack surfaces (7):

\begin{itemize}[leftmargin=*]
\item \textbf{Hidden/invisible text}: white-on-white text runs, font-size-zero text, and zero-width Unicode characters (U+200B, U+200C, U+200D, U+FEFF) are stripped from the extracted text layer.
\item \textbf{Embedded metadata}: PDF (Portable Document Format)/DOCX (Office Open XML Document) author, company, and comment metadata fields are cleared.
\item \textbf{Non-visible layers}: OCR (Optical Character Recognition)/text layers that do not match the rendered visual content are flagged for manual inspection rather than silently discarded (a mismatch is itself a signal of tampering).
\item \textbf{Macros/embedded scripts}: any embedded executable content is stripped; presence is logged as a rejection-worthy integrity flag.
\end{itemize}

Output: a plain-text/Markdown canonical version of the manuscript, plus a sanitization log (what was removed, where) retained for audit but never shown to reviewers. Implementation: a custom Node.js pipeline (\texttt{pdf-parse}/\texttt{mammoth} for extraction plus regex/Unicode-category stripping), avoiding dependence on a third-party sanitization SaaS (Software as a Service).

\subsection{Anonymization protocol}

Applied to the sanitized canonical text before it reaches either LLMs or human reviewers:

\begin{table}[htbp]
\centering
\small
\begin{tabular}{@{}p{0.32\textwidth}p{0.58\textwidth}@{}}
\toprule
\textbf{Element} & \textbf{Action} \\
\midrule
Author name(s) & Removed from title page, running headers, footers \\
Affiliation(s) & Removed \\
Acknowledgments section & Removed (frequently reveals funding source/institution) \\
Email / ORCID (Open Researcher and Contributor ID) & Removed \\
Self-citations & Flagged, not removed --- replaced with a generic marker (e.g., ``[AUTHOR, year]'') so reviewers can assess citation practice without identifying the author via citation pattern \\
Distinctive first-person narrative (``in our previous work...'') & Flagged for manual editor review --- cannot be reliably auto-redacted without damaging scientific content \\
\bottomrule
\end{tabular}
\end{table}

Custody of the original, non-anonymized manuscript is held by the anonymization system itself --- not by the editor --- as an encrypted, access-controlled record keyed to the manuscript's anonymized ID. This record remains locked through stages 2--5 until the editor's publication decision is made, and unlocks only if that decision is an approval --- never as a byproduct of merely reaching stage 5 or of the AI and human layers having passed. The editor gains access to it only at the moment of, and because of, an approval decision, keeping the ``editor does not know author identity'' property structural rather than a matter of discretion. Neither the LLMs nor the human reviewers ever receive this record.

\subsection{LLM pre-screening rubric}

Each of the three LLM reviewers receives an identical, versioned prompt containing the anonymized manuscript and an explicit rubric with named criteria, a fixed numeric scale, and level descriptors per criterion --- not an open-ended ``review this paper'' instruction, since unstructured prompting is what produces the miscalibrated, overconfident scoring documented in prior work (13).

Proposed initial rubric (four criteria, 1--5 scale each):

\begin{table}[htbp]
\centering
\small
\begin{tabular}{@{}p{0.19\textwidth}p{0.25\textwidth}p{0.25\textwidth}p{0.22\textwidth}@{}}
\toprule
\textbf{Criterion} & \textbf{1 (Reject)} & \textbf{3 (Borderline)} & \textbf{5 (Strong)} \\
\midrule
Methodological rigor & Study design cannot support stated conclusions & Design is adequate but has notable gaps & Design is sound and well-justified \\
Statistical coherence & Tests/analyses mismatched to data or design & Broadly appropriate, minor concerns & Fully appropriate, correctly reported \\
Clarity \& structure & Findings cannot be followed without external context & Understandable with effort & Clear, well-organized, self-contained \\
Citation integrity (superficial) & Citations unverifiable/absent, or claims uncited & Mostly verifiable, some gaps & Fully verifiable, well-supported \\
\bottomrule
\end{tabular}
\end{table}

Each LLM must also output structured chain-of-thought justification per criterion before the numeric score. Aggregation rule (working value, to be recalibrated once pilot data exists): mean score $\geq$3.5 across all three LLMs and no single criterion below 2 $\rightarrow$ pass to human review; otherwise $\rightarrow$ return to author with the three LLMs' structured feedback merged (see below for the additional, stricter two-of-three condition that gates the opt-in AI-only-reject path). Here, ``no single criterion below 2'' refers to that criterion's mean across the three LLMs, consistent with how the overall $\geq$3.5 threshold is itself a cross-LLM mean; this is distinct from the AI-only-reject condition below, which by design operates on individual LLM scores rather than a per-criterion mean.

\subsection{Prompt versioning \& reproducibility}

For every review cycle, the system logs and permanently stores: exact prompt text (including rubric version), model identifier and checkpoint/build version, temperature and other sampling parameters, and execution timestamp --- addressing the reproducibility risk of an undetected, provider-side silent model update between two reviews.

\subsection{Anti-prompt-injection safeguard (defense in depth)}

Sanitization removes known injection vectors before the text reaches an LLM; as a second layer, the review prompt instructs each LLM to treat the manuscript strictly as data, never as instructions, and to flag --- rather than obey --- any embedded text attempting to direct its output. Any such flag automatically fails the Check gate and routes to manual editor inspection, regardless of the scores produced.

\subsection{LLM $\rightarrow$ human escalation criteria, and the author-authorized rejection mode}

At submission, the author chooses one of two escalation policies for their manuscript --- a consent decision made in advance, not a per-case option the system applies opportunistically:

\begin{itemize}[leftmargin=*]
\item \textbf{``Always human review''} (default): the LLM layer can only pass a manuscript forward or request revision. It can never issue a terminal rejection --- every manuscript that is not withdrawn by the author eventually reaches a human reviewer or editor.
\item \textbf{``AI-authorized rejection''} (opt-in): the author consents, in advance, to let the three LLMs jointly reject the manuscript when \textbf{two or more of the three} independently score below the pass threshold (3.5) \textbf{on the same criterion} --- a deliberate two-of-three bar, not three-of-three, so a single model having an off day cannot trigger a rejection alone, while genuine, convergent weakness still can. When triggered, the manuscript is returned for revision without any human ever having seen it, along with the specific low-scoring criteria and the LLMs' justification --- an actionable error filter, not a bare verdict. Because no human peer or editor is exposed to the submission, there is no reviewer bias against the author for a future resubmission, and no social cost of a ``visible'' rejection. This is available only because the author affirmatively opted in; it is never the default.
\end{itemize}

Conditions are evaluated in the order listed below --- the AI-only-reject condition is checked first and, if triggered, supersedes Pass regardless of overall mean, since Pass and AI-only-reject are not mutually exclusive as scored (a manuscript can clear the $\geq$3.5 overall mean while still having two of three LLMs converge below 3.5 on one specific criterion).

\begin{table}[htbp]
\centering
\small
\begin{tabular}{@{}p{0.20\textwidth}p{0.48\textwidth}p{0.24\textwidth}@{}}
\toprule
\textbf{Outcome} & \textbf{Condition} & \textbf{Routing} \\
\midrule
Integrity flag & Injection attempt or sanitization mismatch detected & $\rightarrow$ editor, bypassing normal flow, regardless of chosen escalation policy \\
AI-only reject \textit{(opt-in policy only, evaluated before Pass)} & Two or more of the three LLMs independently score below 3.5 on the same criterion & $\rightarrow$ returned to author with the specific low-scoring criteria and LLM justification, as private optimization feedback; no human record created \\
Pass & Not an AI-only-reject or integrity-flag case; mean score $\geq$3.5 across all three LLMs, no per-criterion mean below 2 & $\rightarrow$ Human Review \\
Revise & Not an AI-only-reject or integrity-flag case; below pass threshold, but fewer than two LLMs score below 3.5 on the same criterion (single-model outlier) & $\rightarrow$ returned to author with merged LLM feedback \\
\bottomrule
\end{tabular}
\end{table}

\textbf{Information visibility across layers.} Passing the Check gate forwards only the anonymized manuscript to the three human reviewers --- never the LLM scores, justifications, or threshold that produced the ``Pass'' outcome. Human reviewers know only that a manuscript cleared AI pre-screening, and render their assessment independently, with no visibility into what the AI layer concluded. This is a deliberate safeguard against automation bias --- the tendency for a prior algorithmic judgment to anchor subsequent human judgment --- and mitigates the score-inflation risk documented in the AI-assisted peer review literature (5,13). The editor is the only actor with visibility into \textbf{both} the AI scores and the human assessments, combining them at stage 5 to inform the final publication decision. De-anonymization is triggered only once the editor approves the manuscript. If not approved, the manuscript remains anonymized and is returned to the author with the consolidated observations from every prior stage.

\subsection{Infrastructure \& data governance}

All three LLM reviewers run on locally-hosted, open-weight infrastructure (candidates: Gemma 3 27B, Qwen3 14B/32B --- both feasible on a single high-end consumer GPU (Graphics Processing Unit) at Q4 (4-bit quantization), or a small multi-GPU setup to run three instances concurrently), ensuring manuscript content never leaves the journal's own infrastructure during AI pre-screening. This is a stronger guarantee than even the best commercial API (Application Programming Interface) terms: Anthropic's standard Messages API, for instance, does not retain conversation content by default, and a Zero Data Retention (ZDR) agreement --- a stronger, contractual version of the same commitment --- is available only to eligible accounts, with retention still possible for up to two years in flagged policy-violation cases even under ZDR (16). Both arrangements still require trusting a third party's infrastructure, retention practices, and continued compliance --- a dependency local hosting removes entirely, regardless of how narrow that third party's own stated retention window already is. Commercial APIs and consumer-tier conversational AI products are explicitly excluded from the pre-screening pipeline; the human peer review stage is unaffected, since it involves no data transmission to any LLM provider. This design is a direct, voluntary response to the confidentiality rationale --- not a quality objection --- underlying NIH's and NSF's bar on reviewers submitting unpublished material to third-party generative AI (8,9).

\subsection{Ethical disclosure}

Consistent with ICMJE's (International Committee of Medical Journal Editors) 2023 AI-disclosure guidance and COPE's (Committee on Publication Ethics) position that AI use in review must be transparently disclosed (17), authors are informed before submission that manuscripts will be processed by local LLMs as part of pre-screening, prior to any human review. Since COPE explicitly prohibits reviewers and editors from using generative AI to produce peer review reports or publication decisions (17), the LLM layer here performs pre-screening triage only --- the substantive peer review report and the publication decision remain exclusively human acts.

\section{Implications, Ethical Positioning, and Open Questions}

\subsection{Navigating COPE's position on AI-generated review reports}

COPE's guidance on artificial intelligence states plainly that peer reviewers and editors must not use generative AI tools to produce peer review reports, editorial assessments, or publication decisions; supportive use is permitted only with transparent disclosure (17). Any architecture in which an LLM materially influences a manuscript's progression before human review must address this directly. We argue the tension is resolvable on three grounds.

First, role separation: the AI layer's output is a rubric-scored triage signal --- structurally analogous to existing non-AI screening tools already normalized in scholarly publishing, such as SciScore or StatReviewer (15) --- not a substitute peer review report. It is never presented to the editor as a human reviewer's assessment, and is logged and labeled as AI-generated triage, kept structurally and visually distinct from the human review stage.

Second, exclusive human authorship: the peer review report is authored solely by the three human reviewers, and the publication decision solely by the human editor. The AI layer's output never appears inside either artifact --- it operates strictly upstream of both.

Third, disclosure as the compliance mechanism, not a courtesy: COPE's core requirement is transparent disclosure, not proximity avoidance. Before submission, authors are informed --- unambiguously and in advance --- of the AI layer's existence, scope, and the escalation policy they selected.

One boundary case deserves explicit treatment: the opt-in ``AI-authorized rejection'' mode, in which the AI layer alone returns a manuscript to the author and no human reviews it in that cycle. Because no human reviewer or editor adopts, endorses, or relies on an AI judgment, this path arguably does not constitute an ``editorial decision'' in COPE's sense --- the manuscript simply never enters the editorial process, comparable to an author privately deciding not to submit yet. This reading is defensible but carries genuine interpretive risk, since it has not been tested against COPE guidance in practice; we recommend any journal adopting this mode state its position explicitly in its own public AI policy rather than relying on inference from COPE's general language.

\subsection{The ethical case for author-authorized, unwitnessed AI rejection}

The opt-in ``AI-authorized rejection'' mode targets a specific, well-documented harm of traditional peer review that has received little architectural attention. Manuscript rejection carries a measurable psychological cost --- authors describe reactions resembling stages of grief, and rejection without adequate social support has been linked to isolation, reduced creativity, and withdrawal from research altogether (18). Qualitative work using photovoice methodology documents coping strategies from avoidance to active reframing, underscoring that rejection is an emotionally consequential event, not merely a procedural outcome (19). This burden is compounded by silence and stigma: researchers overwhelmingly share acceptances but conceal rejections, and journals' own reasoning reinforces this --- an analysis of 304 rejection letters found over 97\% attributed rejection to the authors' own shortcomings, under 3\% acknowledged journal-side limitations, and, among 266 desk-rejection letters specifically, 40.6\% cited only a generic ``did not meet the journal's criteria'' and a further 18.8\% cited only a generic ``strict evaluation process'' --- together, close to 60\% giving no substantive, actionable reason (20). This pattern discourages authors from recognizing rejection as a near-universal, systemic experience rather than a personal failing. The burden also has an anticipatory dimension: rejection is described in the literature as ``a dreaded fear that most authors anticipate'' (21). We suspect --- though we are not aware of direct empirical evidence on this specific point, and flag it as a hypothesis rather than an established finding --- that for some authors this dread extends upstream, discouraging manuscript completion or submission altogether. If so, a private, unwitnessed pre-check would lower the barrier to attempting submission, not only soften an eventual rejection's impact.

We highlight one structural feature common to this documented harm: the rejection is witnessed. An editor or reviewer sees the manuscript, forms a judgment, and --- in the author's perception --- now associates the author with that judgment. The ``AI-authorized rejection'' pathway isolates the diagnostic value of an early rejection from its socially witnessed component, by making the outcome, when authorized in advance by the author, visible to no human reviewer or editor at all. We frame this explicitly as a testable hypothesis for the pilot phase, not an established benefit: we are not aware of direct empirical evidence that unwitnessed AI rejection reduces the psychological cost of rejection relative to human-witnessed rejection, only literature suggesting the witnessed, socially-recorded nature of rejection is central to why it produces shame and silence (20). Validating this claim, rather than assuming it, is an explicit item for future work.

This design must also answer a legitimate objection from the automation-bias literature: assuming the absence of human oversight is safe simply because no human judgment is being second-guessed is exactly the assumption two decades of automation-bias scholarship warns against (22). An unaudited AI-only path could, in principle, systematically and silently reject legitimate work --- for instance, due to stylistic bias or unfamiliarity with a minority subfield. We do not consider this fully resolved by opt-in consent alone, and propose three mitigations: (i) the mode is strictly opt-in, never default, and withdrawable by the author for any future submission; (ii) every AI-only rejection is logged in full (rubric scores, model versions, justification), enabling periodic aggregate auditing for systematic bias without reintroducing a human witness to any individual rejection; and (iii) authors are informed, as part of meaningful consent rather than a procedural checkbox, of exactly what this trade-off is.

\subsection{The speed argument, quantified}

The dominant bottleneck in traditional peer review is not reviewer deliberation but reviewer recruitment. Editors needed to send an average of 2.7 reviewer invitations for every accepted review in 2017 --- a figure projected to rise to 3.6 by 2025 as reviewer responsiveness declines --- since more than half of all review invitations were already being declined that year (23). This recruitment delay compounds with the review-writing period itself to produce the roughly 13-week (91-day) average time to first decision observed across fields (11).

Our architecture targets this bottleneck rather than claiming AI reviews ``better'' or ``faster'' than humans. Because the AI layer operates in seconds to low minutes per manuscript (24), manuscripts it scores as not yet ready are returned to the author before ever entering the multi-week human-reviewer-recruitment funnel; only manuscripts scored ready proceed to recruiting human reviewers at all. This is consistent with existing evidence that front-loading triage ahead of reviewer recruitment measurably compresses time-to-decision even when performed by editors rather than AI: immediate (desk) rejections in medicine average around 11 days, with 62\% resolved within one week, and only 8\% taking more than four (11). We expect an AI-based triage layer to compress this stage further. We deliberately do not promise a fixed turnaround: the defensible claim is a reduction in how many manuscripts ever enter the slowest stage of the pipeline, not a guarantee about review depth, rigor, or a specific number of days.

\subsection{Comparison with existing systems and precedents}

The table below situates our proposal against the landscape of related approaches, spanning ungoverned status-quo AI use, narrow-scope commercial screening tools, research benchmarks, and the first large-scale institutional deployment of disclosed AI-assisted review.

\begin{landscape}
\begin{table}[p]
\centering
\scriptsize
\begin{tabular}{@{}p{0.16\textwidth}p{0.17\textwidth}p{0.13\textwidth}p{0.16\textwidth}p{0.13\textwidth}p{0.15\textwidth}p{0.17\textwidth}@{}}
\toprule
\textbf{System} & \textbf{Scope} & \textbf{Independent AI reviewers} & \textbf{Human final decision} & \textbf{Blind protocol} & \textbf{Local-hosting / confidentiality-first} & \textbf{Author consent over AI role} \\
\midrule
Traditional peer review & General merit review & 0 & Yes (only option) & Single/double-blind & N/A & N/A \\
Undisclosed AI use (e.g., ICLR 2026, $\sim$21\% of reviews fully AI-generated) (4) & Whatever the reviewer delegates & Unknown, unaudited & Nominally, but reviewer's own judgment is bypassed without disclosure & Nominal only --- undermined by non-disclosure & No & No --- this is precisely the undisclosed status quo our design opposes \\
SciScore, StatReviewer, Penelope.ai, UNSILO, RobotReviewer (15) & Narrow: methods transparency, statistics, journal-requirement compliance, concept extraction, RCT (Randomized Controlled Trial) bias & 1 tool, narrow task & Yes, fully separate from tool output & Unchanged & No & Disclosed as editorial tooling, but no consent choice over role \\
Shen et al. (2026), JMIR (Journal of Medical Internet Research) AI (10) & Research benchmark: single-model quartile classification & 1 model per test run & N/A (research setting, not deployed) & N/A & No & N/A \\
ICLR 2025 Review Feedback Agent (14) & Feedback to human reviewers, not independent assessment & 5-LLM pipeline, but output feeds humans, not decisions & Yes & Unchanged & No & Disclosed to reviewers \\
AAAI-26 (Association for the Advancement of Artificial Intelligence) AI Review Pilot (25) & General technical review at conference scale (22,977 papers) & 1 AI review per paper & Yes --- AI output excluded numeric scores and accept/reject recommendations by design & Unchanged & Not addressed & Disclosed to authors/PC (Program Committee), no consent choice over role \\
\textbf{Our proposal} & General scientific merit pre-screening & 3 independent LLMs & Yes, mandatory (opt-in exception; see Section~\ref{sec:architecture}) & Triple-blind (novel layer added ahead of existing human blind stages) & Yes --- all three models locally-hosted by design & Yes --- author selects escalation policy at submission, including whether AI may return manuscripts without human involvement \\
\bottomrule
\end{tabular}
\end{table}
\end{landscape}

The AAAI-26 pilot (25) is the strongest available evidence that disclosed, large-scale AI-assisted review is institutionally acceptable: all 22,977 main-track papers were each assigned exactly one AI-generated review, clearly flagged as such, produced in under a day at a reported cost below \$1 per paper and achieving an average recall gain of +0.21 over baseline ($p = 6.1\times10^{-30}$). When surveyed, authors and program-committee members rated these AI-written reviews above their human counterparts on measures including technical accuracy. Critically, the pilot's own safeguards mirror the role-separation principle we adopt above: AI output carried no numeric score and no accept/reject recommendation, and did not replace any human reviewer. Our proposal differs by using three independent models rather than one (reducing single-model idiosyncrasy, consistent with the instability documented in (13)), by embedding this inside a triple-blind protocol rather than a single-blind review, by requiring local hosting for every AI review layer as a fixed guarantee rather than an undisclosed infrastructure choice, and by making the escalation policy an author-facing design choice.

\section{Limitations}

\textbf{No empirical validation yet.} This architecture has not been piloted on real manuscripts. The rubric thresholds proposed in Section~\ref{sec:architecture} (e.g., mean score $\geq$3.5, no criterion below 2) are working values, not calibrated parameters, requiring validation against a pilot comparing AI pre-screening outcomes to final human/editorial decisions. Any claims here about accuracy, false-positive/false-negative rates, or realized time savings are projections from the literature and system design, not measured outcomes.

\textbf{Residual model bias and instability.} Even with an explicit rubric, LLM-generated assessments are systematically less variable, more positive, and overconfident relative to human reviewers (13), and have been documented to inflate acceptance-relevant scores at scale (5). Three independent models mitigate single-model idiosyncrasy but do not eliminate correlated bias --- open-weight models in comparable ranges often train on overlapping data and may share blind spots, such as underweighting genuine methodological novelty, a documented weakness relative to fact-checking and literature-coverage tasks (26).

\textbf{Capability gap between locally-hosted and frontier models.} Confidentiality-first design requires all three reviewers to run on locally-hosted, open-weight models (14B--32B parameter range), with no frontier commercial model in the loop at any stage of AI pre-screening. Such models remain a step behind frontier commercial models (e.g., Claude Opus, GPT-5.x) on subtle reasoning tasks --- a trade-off accepted deliberately in exchange for eliminating third-party data exposure entirely, and precisely why human peer review remains the mandatory final layer, rather than an optional safeguard.

\textbf{Sanitization is necessary but not sufficient.} The sanitization protocol targets known prompt-injection vectors, but adversarial techniques evolve; documented cases so far involve relatively unsophisticated methods (white-on-white text) (7), and we should not assume this remains true. The in-prompt instruction to treat manuscript content strictly as data is a second layer of defense, not a guarantee.

\textbf{Scale and cost were not evaluated here.} Unlike the AAAI-26 pilot, which reported concrete per-paper cost and processing-time figures at conference scale (25), we have not yet measured the compute cost or throughput of running three independent, locally-hosted models per submission at the volume a full-scale journal deployment would require. This is deferred to the pilot phase.

\textbf{The ``AI-authorized rejection'' opt-in mode is ethically novel and untested.} This mode's compliance posture is a reasoned interpretation, not a tested or externally validated one, and depends on continued author autonomy and genuinely informed consent rather than default-on convenience.

\textbf{Regulatory posture is voluntary, not binding.} Alignment with NIH/NSF confidentiality rationale and COPE disclosure norms is a deliberate ethical choice by an independent journal outside those agencies' jurisdiction, not a compliance requirement, and should not be overstated as regulatory endorsement.

\section{Conclusion}

Peer review is already being reshaped by large language models, largely through undisclosed use: roughly one in five reviews at a major 2026 machine learning conference were AI-generated, and over a hundred hallucinated citations reached an accepted proceedings (4,6). The question is no longer whether AI will touch peer review, but whether that involvement is governed and disclosed, or left to accumulate silently.

We have described an alternative: a triple-blind architecture in which three independent, locally-hosted LLMs perform structured, rubric-based pre-screening ahead of mandatory human peer review and editorial decision-making, voluntarily satisfying the confidentiality rationale behind NIH and NSF's restrictions on generative AI in review (8,9). The design keeps the review report and the publication decision exclusively human, as COPE requires (17), while reducing how many manuscripts enter the slowest stage of the traditional process: reviewer recruitment. It also gives authors, not just the journal, a documented choice over how AI may be involved in their submission, including an opt-in mode addressing the specific, evidenced harm of witnessed manuscript rejection (18,20) --- a hypothesis for testing, not a settled claim.

This is architecture, not yet evidence, and validation should happen in two stages. First, a preparatory bench validation: running the full pipeline --- sanitization, AI review, and the Check gate --- on labeled corpora of accepted, rejected, and prompt-injection-seeded manuscripts, to test whether the gate's routing converges on the known ground truth. Unlike Shen et al., this would validate the whole pipeline, not a single classification task, and would be the first proposed test of an AI pre-screening layer's robustness against adversarial manuscripts. Two caveats apply: thresholds should be calibrated on a held-out set to avoid overfitting, and genuinely rejected manuscripts are hard to source --- rejection is rarely shared, for the reasons of stigma and silence discussed above (20) --- making deliberately flawed manuscripts (12) a more tractable substitute.

Second, a live three-arm pilot: (i) AI-only review, (ii) our hybrid architecture, and (iii) conventional human-only review, compared on decision quality, time-to-decision, and psychological burden. This directly tests the paper's central claim: if the hybrid arm does not outperform AI-only on decision quality, or does not reduce time-to-decision relative to conventional review, the architecture's value proposition needs revisiting. We see disclosed, human-supervised, multi-model pre-screening as a template other journals under the same pressure can adapt --- the real choice is not whether AI touches peer review, but whether its presence is governed or left unacknowledged.

\section*{Acknowledgements}

During the preparation of this work, a generative AI tool was only used for language editing and refinement. The tool was not used to generate the intellectual content, arguments, or conclusions.

\section*{Footnote}

\textbf{Reporting Checklist:} Not applicable --- no specific EQUATOR reporting guideline applies to this architecture-proposal manuscript, which does not report a clinical trial, systematic review, or empirical dataset.

\textbf{Data Sharing Statement:} No datasets were generated or analyzed for this manuscript. The system architecture described is a design proposal that has not yet been piloted on real manuscripts (see Limitations).

\textbf{Funding:} None.

\textbf{Conflicts of Interest:} The author has completed the ICMJE uniform disclosure form. The author has no conflicts of interest to declare.

\textbf{Ethical Statement:} The authors are accountable for all aspects of the work in ensuring that questions related to the accuracy or integrity of any part of the work are appropriately investigated and resolved.

\textbf{Open Access Statement:} This is an Open Access article distributed in accordance with the Creative Commons Attribution 4.0 International License (CC BY 4.0), which permits unrestricted use, distribution, and reproduction in any medium, provided the original work is properly cited. See: \url{https://creativecommons.org/licenses/by/4.0/}.

\section*{Abbreviations}

\begin{description}[leftmargin=2.2cm, style=nextline]
\item[AAAI] Association for the Advancement of Artificial Intelligence
\item[AI] artificial intelligence
\item[API] application programming interface
\item[COPE] Committee on Publication Ethics
\item[DOCX] Office Open XML Document
\item[GPU] graphics processing unit
\item[ICLR] International Conference on Learning Representations
\item[ICMJE] International Committee of Medical Journal Editors
\item[ICML] International Conference on Machine Learning
\item[JMIR] Journal of Medical Internet Research
\item[LLM] large language model
\item[NeurIPS] Conference on Neural Information Processing Systems
\item[NIH] National Institutes of Health
\item[NSF] National Science Foundation
\item[OCR] optical character recognition
\item[ORCID] Open Researcher and Contributor ID
\item[PC] program committee
\item[PDF] Portable Document Format
\item[RAG] retrieval-augmented generation
\item[RCT] randomized controlled trial
\item[SaaS] software as a service
\item[ZDR] zero data retention
\end{description}

\section*{References}

\begin{enumerate}[leftmargin=*, label={\arabic*.}]
\item Neural Information Processing Systems Foundation. NeurIPS 2025 fact sheet [Internet]. 2025 [cited 2026 Jul 6]. 21,575 submissions, 5,290 accepted. Available from: \url{https://media.neurips.cc/Conferences/NeurIPS2025/press/NeurIPS2025-Fact_Sheet.pdf}
\item International Conference on Learning Representations. ICLR 2025 fact sheet [Internet]. 2025 [cited 2026 Jul 6]. 11,603 submissions, 3,704 accepted (32\%). Available from: \url{https://media.iclr.cc/Conferences/ICLR2025/ICLR2025_Fact_Sheet.pdf}
\item International Conference on Machine Learning. ICML 2025 statistics --- 12,107 submissions, 3,260 accepted (26.93\%) [Internet]. Paper Copilot; 2025 [cited 2026 Jul 6]. Available from: \url{https://papercopilot.com/statistics/icml-statistics/}
\item Emi B. Pangram predicts 21\% of ICLR reviews are AI-generated [Internet]. Pangram Labs; 2025 Nov 18 [cited 2026 Jul 6]. Available from: \url{https://www.pangram.com/blog/pangram-predicts-21-of-iclr-reviews-are-ai-generated}
\item Russo et al. The AI Review Lottery: Widespread AI-Assisted Peer Reviews Boost Paper Scores and Acceptance Rates. arXiv:2405.02150 / DOI 10.1145/3757667 (PACMHCI/ACM CSCW 2025)
\item GPTZero. GPTZero finds 100 new hallucinations in NeurIPS 2025 accepted papers [Internet]. 2026 Jan [cited 2026 Jul 6]. Available from: \url{https://gptzero.me/news/neurips/}
\item Lin Z. Hidden Prompts in Manuscripts Exploit AI-Assisted Peer Review. arXiv:2507.06185 (Jul 2025)
\item National Institutes of Health. NOT-OD-23-149: The use of generative artificial intelligence technologies is prohibited for the NIH peer review process [Internet]. 2023 Jun 23 [cited 2026 Jul 6]. Available from: \url{https://grants.nih.gov/grants/guide/notice-files/NOT-OD-23-149.html}
\item National Science Foundation. Notice to the research community: use of generative artificial intelligence technology in the NSF merit review process [Internet]. 2023 Dec [cited 2026 Jul 6]. Available from: \url{https://www.nsf.gov/news/notice-to-the-research-community-on-ai}
\item Shen SM, et al. Evaluation of Large Language Models for Peer Review in Transplantation Research: Algorithm Validation Study. JMIR AI 2026;5:e84322. PMID 41672474
\item Huisman J, Smits J. Duration and quality of the peer review process: the author's perspective. Scientometrics. 2017;113(1):633-650. DOI: 10.1007/s11192-017-2310-5
\item Liu Y, Shah NB. ReviewerGPT? An Exploratory Study on Using Large Language Models for Paper Reviewing. arXiv:2306.00622 (Jun 2023)
\item Żurawicki K, Farganus J, Gaweł A, Bystroński M, Kajdanowicz TJ. PRAIB: Peer Review AI Benchmark of Behaviour of LLM-Assisted Reviewing. arXiv:2605.29815
\item Thakkar N, Yuksekgonul M, Silberg J, Garg A, Peng N, Sha F, Yu R, Vondrick C, Zou J. Can LLM feedback enhance review quality? A randomized study of 20K reviews at ICLR 2025. arXiv:2504.09737
\item Kousha K, Thelwall M. Artificial intelligence to support publishing and peer review: A summary and review. Learned Publishing. 2024;37:4-12. DOI: 10.1002/leap.1570
\item Anthropic. API and data retention [Internet]. 2026 [cited 2026 Jul 6]. Available from: \url{https://platform.claude.com/docs/en/manage-claude/api-and-data-retention}
\item Committee on Publication Ethics. Artificial intelligence (AI) in decision making [Internet]. [cited 2026 Jul 6]. Available from: \url{https://publicationethics.org/resources/discussion-documents/ai-artifical-intelligence-decision-making}; COPE Focus on Artificial Intelligence. Available from: \url{https://publicationethics.org/cope-focus/artificial-intelligence}
\item Day NE. The Silent Majority: Manuscript Rejection and Its Impact on Scholars. Academy of Management Learning \& Education. 2011;10(4):704-718. DOI 10.5465/amle.2010.0027
\item Argan M, Dinç H, Tokay Argan M, Özer A. What does rejection look like? A photovoice study on emotions and coping regarding manuscript rejection. Current Psychology. 2023. PMC9842215
\item Nguyen M-H, Vuong Q-H. The Unfair Burden of Rejection on Researchers: Transitioning From Editors as Gatekeepers to Facilitators of Knowledge Production. Learned Publishing. 2025;38(4):e2027. DOI: 10.1002/leap.2027
\item Ali MJ. The science and philosophy of manuscript rejection. Indian Journal of Ophthalmology. 2021. DOI 10.4103/ijo.IJO\_1097\_21
\item Pham T. Ethical and legal considerations in healthcare AI: innovation and policy for safe and fair use. Royal Society Open Science. 2025;12(5):241873. DOI: 10.1098/rsos.241873. PMC12076083
\item Publons/Clarivate. Global State of Peer Review Report (inaugural edition, 2017 data) [Internet]. 2018 [cited 2026 Jul 6]. Available from: \url{https://publons.com/static/Publons-Global-State-Of-Peer-Review-2018.pdf}
\item Haryanto CY. LLAssist: Simple Tools for Automating Literature Review Using Large Language Models. arXiv:2407.13993
\item Biswas J, Schoepp S, Vasan G, et al. AI-Assisted Peer Review at Scale: The AAAI-26 AI Review Pilot. arXiv:2604.13940 (Apr 2026)
\item ReviewerToo (Sahu G, Larochelle H, Charlin L, Pal C). arXiv:2510.08867
\item Neural Information Processing Systems Foundation. NeurIPS 2024 fact sheet [Internet]. 2024 [cited 2026 Jul 6]. 17,491 submissions, 4,497 accepted. Available from: \url{https://media.neurips.cc/Conferences/NeurIPS2024/NeurIPS2024-Fact_Sheet.pdf}
\item International Conference on Learning Representations. ICLR 2024 fact sheet [Internet]. 2024 [cited 2026 Jul 6]. 7,262 submissions, 2,260 accepted (31.1\%). Available from: \url{https://media.iclr.cc/Conferences/ICLR2024/ICLR2024_Fact_Sheet.pdf}
\item ICML Conference (@icmlconf). ICML 2026 reaching 24,371 submissions [Internet]. X (formerly Twitter); 2026 Jan 29 [cited 2026 Jul 6].
\end{enumerate}

\end{document}